\documentclass[runningheads]{llncs}

\usepackage[utf8]{inputenc}

\usepackage{hyperref}
\usepackage{graphicx}
\usepackage{cite}
\usepackage{xcolor}

\usepackage[english]{babel}

\usepackage{amsfonts, amsmath, amssymb}
\usepackage{indentfirst}
\usepackage[all]{xy}
\usepackage{algpseudocode}
\usepackage{algorithm}

\algblock{Input}{EndInput}
\algnotext{EndInput}
\algblock{Output}{EndOutput}
\algnotext{EndOutput}

\DeclareMathOperator{\E}{\mathbb{E}}

\title{On Probability Shaping for 5G MIMO Wireless Channel with Realistic LDPC Codes}
\titlerunning{Probability Shaping of Quadrature Amplitude Modulation}

\author{Evgeny~Bobrov\inst{1, 2, \star}\orcidID{0000-0002-2584-6649 }, Adyan~Dordzhiev\inst{3, \star}}

\institute{Moscow Research Center, Huawei Technologies, Russia \\
\and 
M. V. Lomonosov Moscow State University, Russia \\
\email{eugenbobrov@ya.ru}
\and
National Research University Higher School of Economics, Russia \\
\email{adyandordzhiev@yandex.ru} \\
$^\star$ Equal contribution
}

\begin{document}

\maketitle

\begin{abstract}


Probability Shaping (PS) is a method to improve a Modulation and Coding Scheme (MCS) in order to increase reliability of data transmission. It is already implemented in some modern radio broadcasting and optic systems, but not yet in wireless communication systems. Here we adapt PS for the 5G wireless protocol, namely, for relatively small transport block size, strict complexity requirements and actual low-density parity-check codes (LDPC). We support our proposal by a numerical experiment results in Sionna simulator, showing 0.6 dB gain of PS based MCS versus commonly used MCS.




\end{abstract}

\section{Introduction}

In the 5G New Radio downlink procedure, the user equipment proposes the serving base station for use in the next signal transmission of the optimal modulation and coding scheme (MCS)~\cite{bobrov2021massive} based on quadrature amplitude modulation (QAM). In order to achieve the capacity of the additive white Gaussian noise (AWGN) channel, the transmit signal must be Gaussian distributed. The use of uniformly distributed QAM symbols with optimal coded modulation (CM) leads to a shaping loss of up to 1.53 dB for high order constellations~\cite{forney1984efficient}. Bit-interleaved coded modulation (BICM) with parallel bit-wise demapping as currently employed in LTE leads to an additional loss. 

\begin{figure}
  \centering
    \begin{minipage}{0.33\textwidth}
    \centering
    Complex plane
    \includegraphics[width=\textwidth]{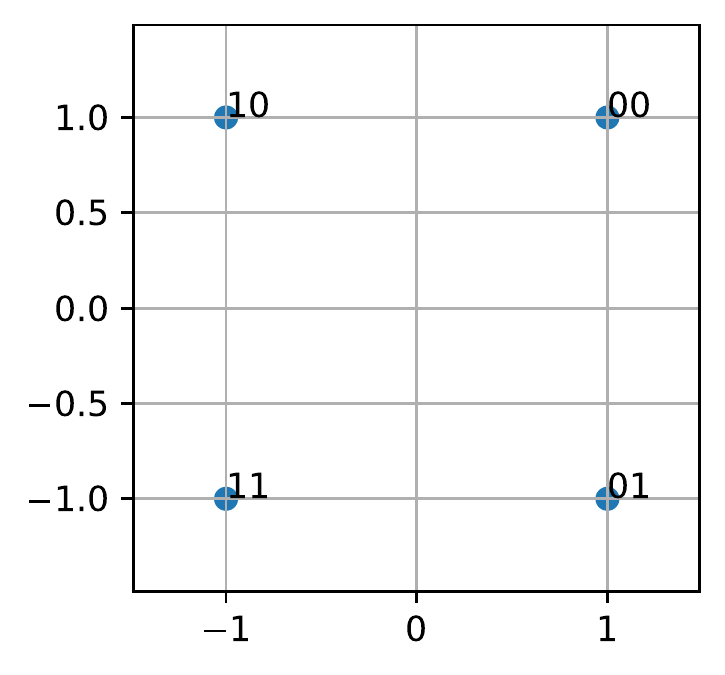}
    QAM-4
  \end{minipage}
  \begin{minipage}{0.31\textwidth}
  \centering
    Complex plane
    \includegraphics[width=\textwidth]{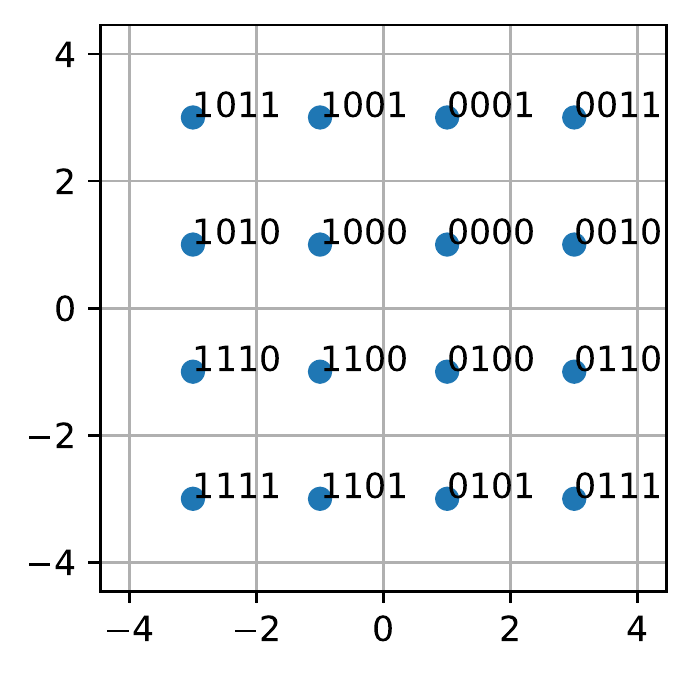}
    QAM-16
  \end{minipage}
  \hfill
  \begin{minipage}{0.33\textwidth}
  \centering
    Complex plane
    \includegraphics[width=\textwidth]{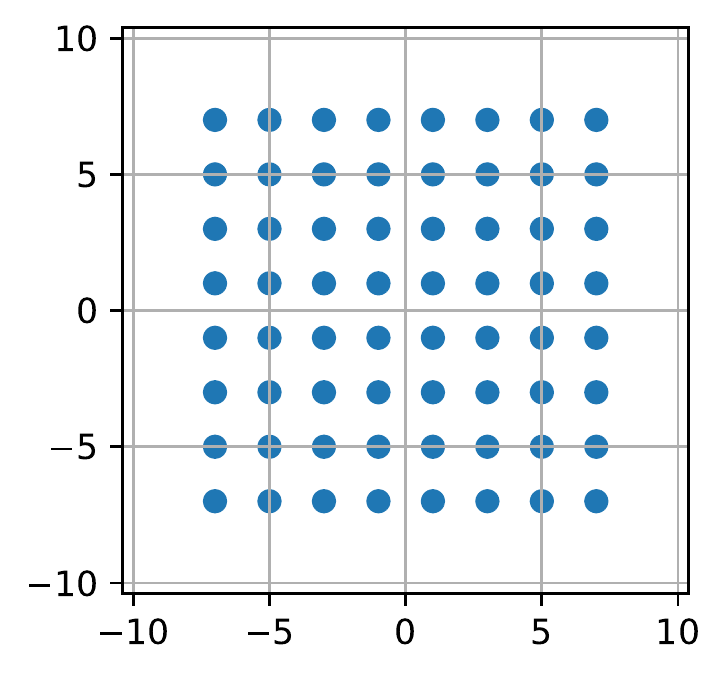}
     QAM-64
  \end{minipage}
    \vfill
  \caption{The base station selects the appropriate QAM scheme for use in the next data transmission. With the increasing of the system quality, the higher QAM can be used.}
    \label{fig:const_plot}
\end{figure}

Non-uniform constellations (NUC) and geometric shaping (GS) have been recently adopted for the next-generation terrestrial broadcast standard~\cite{loghin2016non}. The QAM constellations are optimized for each target signal-to-noise ratio (SNR) by maximizing the BICM capacity for uniformly distributed bits. Note that, in contrast to standard Gray-labeled QAM (Fig.~\ref{fig:const_plot}), the Non-Uniform constellations do not allow for a simple independent demapping of the real and imaginary part. Therefore, one-dimensional NUCs for each real dimension were also studied in~\cite{loghin2016non}, which provide a reduced shaping gain. The performance of BICM can be improved by using non-uniform constellations (NUC), but there remains a gap to the capacity with Gaussian transmit signal.

In the traditional approach of data transmission, each point in a particular constellation has an equal chance of being transmitted. While this technique gives the highest bit rate for a given constellation size, it ignores the energy cost of the individual constellation points. So, as an alternative to GS, it is also possible to adjust the probabilities of the constellation points such that they follow an approximate discrete Gaussian distribution, using the probability shaping (PS) method~\cite{kschischang1993optimal}. Probabilistically shaped coded modulation (PSCM) enables the BICM system to close the gap to the capacity with Gaussian transmit signal. PS is a CM strategy that combines constellation shaping and channel coding.




In the literature, Gallager's error exponent approach has been used to study the achievable information rates of PS~\cite[Ch.~5]{gallager1968information}. In particular, it was shown that the PS method has achievable capacities for additive white Gaussian noise channels~\cite{bocherer2017achievable}. In~\cite{gultekin2020achievable}, the authors revisit the capacity achieving property of PS. The concept of selecting constellation points using a nonuniform Maxwell-Boltzmann PS is investigated in the study~\cite{kschischang1993optimal}. Nonuniform PS signaling scheme reduces the entropy of the transmitter output and, as a result, the average bit rate. However, if low-energy points are picked more frequently than high-energy points, energy savings may (more than) compensate for the bit rate reduction. Authors of~\cite{sillekens2018simple} proposed a new PS distribution that outperforms Maxwell-Boltzmann is studied for the nonlinear fiber channel. In~\cite{idler2017field} the authors successfully tested the suitability of PS constellations in a German nationwide fiber ring of Deutsche Telekom's R\&D field test network. In~\cite{buchali2016rate} the PS method is implemented in 64-QAM coherent optical transmission system. In~\cite{icscan2019probabilistic}, a proposed extension to the 5G New Radio polar coding chain is the introduction of a shaping encoder in front of the polar encoder, which will improve the performance with higher order modulation using this PS scheme.

The main objectives of the study:
\begin{itemize}
    \item In this paper, we investigate the PS Enumerative Sphere Shaping (ESS)~\cite{gultekin2020probabilistic} method known in the literature with respect to a realistic MIMO OFDM wireless channel with LDPC at a given coderate.
    \item We provide numerical experiments on the modern Sionna~\cite{hoydis2022sionna} simulation platform and find local optimal parameters for the ESS method, minimizing BLER and providing a gain of up to 0.6 dB over the QAM-16 baseline.
    \item This study could be interesting from a scientific point of view, since there are almost no published papers on PS that consider such realistic and contemporary scenarios, while considering only theoretical distributions~\cite{neskorniuk2022model}.
\end{itemize}

\begin{figure}
    \centering
    \includegraphics[width=\linewidth]{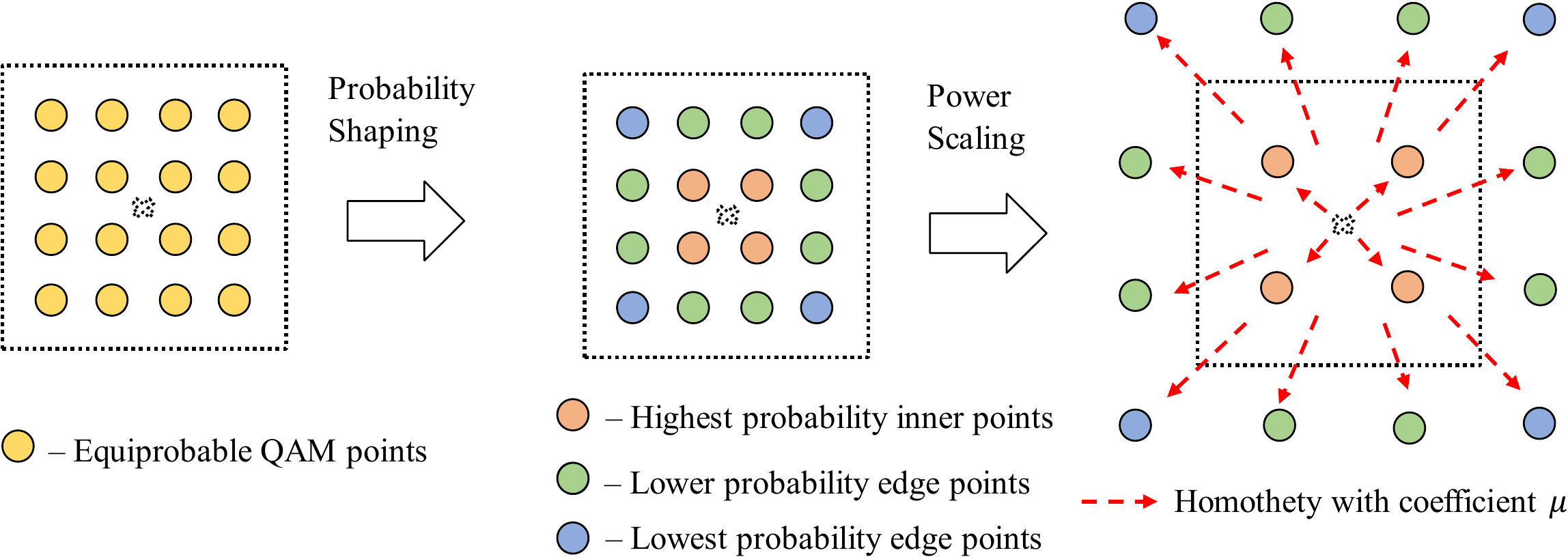}
    \caption{The probability shaping method increases system performance by scaling constellation points, which is allowed while preserving the constellation power ~--- the mathematical expectation of the modulus of the complex points.}
    \label{fig:ps_shaping_scaling}
\end{figure}

The basic principle of PS method is presented in Fig.~\ref{fig:ps_shaping_scaling}. We change the probability of constellation points, which allows us to scale their coordinate with preserving of the mathematical expectation of constellation power.

\section{System Model}
\label{sec:system_model}

A block diagram of the proposed PSCM transmitter and receiver is shown in Fig.~\ref{fig:pscm}. The main difference to conventional BICM is the distribution matcher that maps the uniformly distributed data bits to bit streams with a desired distribution, which determine the amplitudes of the transmitted QAM symbols. The forward error correction (FEC) encoder generates additional parity bits, which are uniformly distributed and determine the signs of the transmitted QAM symbols. This results in an approximately Gaussian distributed transmit signal using the same constellation mapping as in Long-Term Evolution (LTE).

At the receiver side, the QAM demapper calculates the bit-wise log-likelihood ratios (LLRs) based on the observed receive signal, taking the non-uniform transmit symbol distribution into account. These LLRs are fed to the FEC decoder as in conventional BICM, and the decoder output is finally mapped back to data bits by the distribution deshaper. Note that both the distribution matcher and deshaper correspond to simple one-to-one mappings, which can be efficiently implemented.

\section{Optimal Distribution for Probability Shaping}

Let $P_{X} = (p_{1}, \dots, p_{m})$ be the vector of probabilities of each constellation point and $X = (x_{1}, \dots, x_{m})$ is its random variable~--- transmitted points on the constellation, where $m$ is a number of the constellation points.

Let $Y$ be a random variable~--- received points:
\begin{equation}\label{eq:system}
    Y = X + N_0, \qquad N_0\sim \mathcal{CN}(0,\,\sigma^{2}),\qquad 0< \sigma^{2} < \infty,
\end{equation}
where $N_0$ is a Gaussian random variable~--- noise of a channel.


The energy of the constellation is equal to the expectation of $|X|^{2}$, i.e. 
\begin{equation*}
    \E[|X|^{2}] = \sum\limits_{i=1}^{m}p_{i} |x_{i}|^{2}.
\end{equation*}
Our goal is to minimize the energy of the constellation to reduce errors in symbols. In this paper, we study the case when the random variable $X$ is distributed on the QAM constellation.

\textbf{Example.} Initial distribution of $X$ is uniform. For instance, the energy of QAM-16 is equal to 10 since $$\E[|X|^{2}] = \dfrac{1}{16} \cdot (4\cdot2 + 8\cdot10 + 4\cdot18) = 10.$$ Now if we change the distribution of $X$ in such a way that
\begin{itemize}
     \item four points with coordinates $(\pm 1, \pm 1)$ have probability 0.125 
    \item eight points with coordinates $(\pm 1, \pm 3), (\pm 3, \pm 1)$ have probability 0.0375
    \item four points of coordinates $(\pm 3, \pm 3)$ have probability 0.05.
\end{itemize}
In this case, the energy will be equal to 7.6 since $$\E[|X|^{2}] = 0.125 \cdot 4\cdot2 + 0.0375 \cdot 8\cdot10 + 0.05 \cdot 4\cdot18 = 7.6,$$ 
and if we shift the points by multiplying them by the square root of ratio of the energies of the constellations $\sqrt{\frac{10}{7.6}}$, then the energy again become equal to 10. It follows that the points of constellation are further apart, and the variance is the same. So, the probability of error are less.

\subsection{Problem statement}

The physical meaning of the problem~\eqref{eq:primal} is to minimize the constellation energy at a fixed constellation entropy. The entropy $H(X)$ means the amount of information transmitted by the constellation, and the energy $\E[|X|^{2}]$ means the power the transmitter has to expend in transmitting the data.

\begin{equation}\label{eq:primal}
  \left\{
    \begin{aligned}
      & \E[|X|^{2}] = \sum_{i=1}^{m}p_{i} \cdot |x_{i}|^{2} \rightarrow  \underset{P_{X}}{\text{min}}\\
      & \sum_{i=1}^{m}p_{i} = 1\\
      & H(X) = -\sum_{i=1}^{m}p_{i} \cdot \log_{2}p_{i} = const
    \end{aligned}
  \right.
\end{equation}

There is no analytical expression for the problem~\eqref{eq:primal}, and so the constellation points are assumed to have a Maxwell-Boltzmann distribution~\eqref{eq:p_solution} since it is close to the optimal distribution~\cite{kschischang1993optimal} and maximizes the entropy of the constellation with a constraint on its energy:
\begin{equation}\label{eq:p_solution}
    \widehat{p}_{i} = \dfrac{e^{-\mu |x_{i}|^{2}}}{\sum_{j=1}^{m}e^{-\mu |x_{j}|^{2}}}, \quad i = 1, \dots, m
\end{equation}

Parameter $\mu = \mu(\hat{p}_1, \dots \hat{p}_m)$~--- is a scaling of constellation points:
\begin{equation}\label{eq:mu_solution_1}
     \sum_{i=1}^{m}p_{i} \cdot |x_{i}|^{2}  = \sum_{i=1}^{m}\hat{p}_{i} \cdot |\mu x_{i}|^{2} \qquad \Longleftrightarrow \qquad  \mu^{2} = \dfrac{\sum_{i=1}^{m}p_{i} \cdot |x_{i}|^{2} }{\sum_{i=1}^{m}\hat{p}_{i} \cdot |x_{i}|^{2} }
\end{equation}
where $p_{i}$ is the uniform distribution, $\hat{p}_{i}$ is the optimal distribution, and $x_{i}$~is the complex coordinates of the points.







\section{Coded Modulation Design for QAM-16}

\begin{figure}
    \centering
    \includegraphics[width=0.8\textwidth]{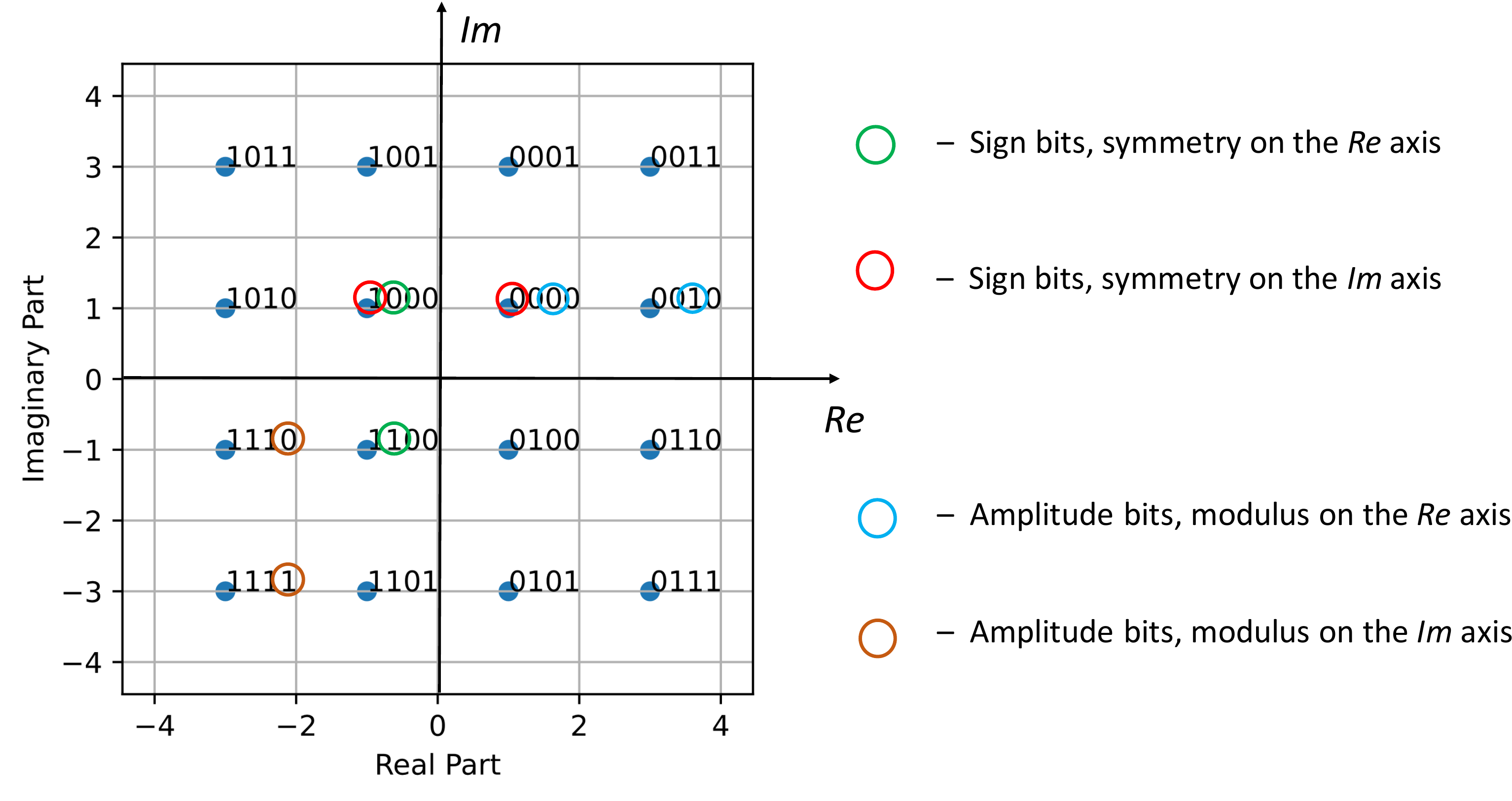}
    \caption{Gray labeling of the sign and amlpitude bits.}
    \label{fig:sign_ampl_bits}
\end{figure}

According to the labelling procedure, we can notice that the first two bits in the binary representation of the constellation points are responsible for symmetry about the coordinate axes, and the last two bits are responsible for absolute value. (Fig.~\ref{fig:sign_ampl_bits}). In what follows, we will refer to the first two bits as sign bits, and the last two bits as amplitude bits. Thus, amplitude bit zero corresponds to points with coordinates~$\pm 1$, and amplitude bit corresponds to points with coordinates~$\pm 3$.


\subsection{Constellation Energy Minimisation}

For the practical finite block-length codes, it is required to implement the Enumerative Sphere Shaping (ESS) method~\cite{gultekin2020probabilistic}.

The energy minimization process is fairly straightforward. We take the constellation points with the smallest absolute value with a higher probability, and the points with the largest absolute value with a lower probability. Thus, we are more interested in constellation points that have more zeros than ones at the amplitude bit positions in the binary representation, and then it is sufficient to maximize the probability of zero at the amplitude bit positions. We also assume that the sign bits are uniformly distributed, i.e. the probability of zero and one of the first two bits in the binary representation of each constellation point is equal to $\frac{1}{2}$.

As noted above, amplitude bit one corresponds to more distant points from the origin, and amplitude bit zero corresponds to closer points. Thus, we can assume that the \emph{energy} of a sequence of $n$ amplitude bits consisting of $k$ ones and $n-k$ zeros, is equal to $$\underbrace {1^{2} + \dotsc + 1^{2}}_{n-k} + \underbrace {3^{2} + \dotsc + 3^{2}}_{k}.$$ It can be seen that the nearest points to the origin have the lowest energy.

For a given number of input amplitude bits $k$ and block length $n$, the most efficient way to change probabilities is to map all possible $2^{k}$ realisations to the $2^{k}$ sequences of $n$ amplitude bits with minimal energy. After that, we can calculate the probability of one $p_{a}(1)$ and probability of zero $p_{a}(0)$  in a set of blocks of length $n$.

Now if we know the distribution of the amplitude bits, then we can find the probability of the constellation points. Each constellation point contains two sign bits and two amplitude bits, so the probability of a point is equal to

\begin{equation}\label{eq: new_distr}
 \widehat{p}_{i} = (\frac{1}{2})^{2} \cdot p_{a} (0) ^{k} \cdot (1 - p_{a}(0))^{1 - k},\quad {i = 1, \dots, 16}
\end{equation}
where $k$ is the number of zero amplitude bits in the bit representation of the constellation point. 

After we have changed the distribution of constellation points, we can calculate the scaling parameter $\mu$ as the ratio of the initial energy to the received energy:
\begin{equation}\label{eq:mu_solution}
   \mu^{2} = \dfrac{\E[|X|^{2}]}{\E[|\widehat{X}|^{2}]} = \dfrac{10}{\sum_{i=1}^{16}\widehat{p}_{i} \cdot |x_{i}|^{2} },
\end{equation}
where $\widehat{X}$ is a new random variable with distribution \ref{eq: new_distr}. Finally, we shift points of the constellation by multiplying them by the parameter $\mu$, thereby reducing the probability of error.


\subsection{Amplitude Shaper and Sign Delay}

\begin{figure}
    \centering
    \includegraphics[width=1\linewidth]{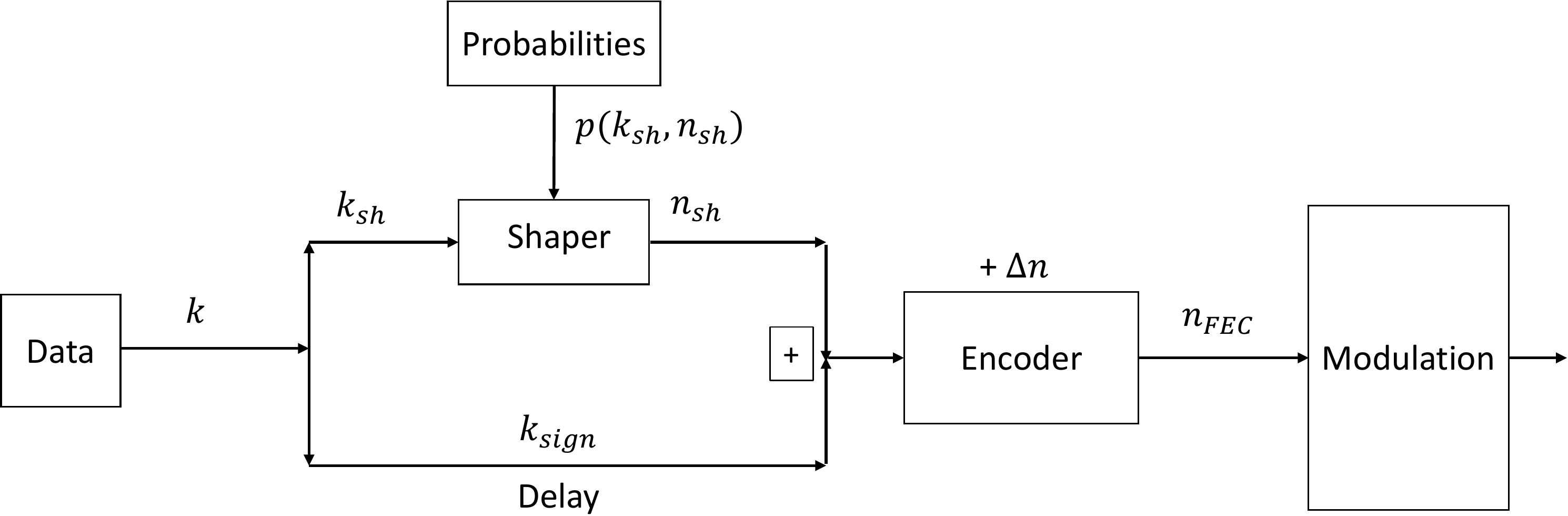}
    \caption{Data flow through the amplitude probability shaper and encoder to modulation.}
    \label{fig:ps_scheme}
\end{figure}

In this subsection we describe the model provided in Fig.~\ref{fig:ps_scheme}. 
Initially, the input is $k$ informational bits with a uniform distribution. These bits are divided into two groups, one of which will be the amplitude bits, and the other group will be part of the sign bits. Amplitude bits are transmitted through the shaper block, which works according to the algorithm described above. The shaper output is a block of a different length, in which the amplitude bits are already distributed according to the algorithm. We will denote the number of bits in the first group by $k_{sh}$,  the number of bits in the second group by $k_{sign}$ and the number of bits at the shaper output by $n_{sh}$.

After that, $k_{sign}$ bits and $n_{sh}$ amplitudes bits are concatenated and encoded using the LDPC procedure. The LDPC procedure, in turn, generates additional $\Delta n$  check bits, which are also considered to be uniformly distributed. We will denote the number of bits at the encoder output as $n_{FEC} = k_{sign} + n_{sh} + \Delta n$. Note that the $k_{sign} + \Delta n$ bits are sign bits, which have a uniform distribution, while the amplitude bits $n_{sh}$ are distributed according to the algorithm. The number of sign $k_{sign}$ and error correction bits $\Delta n$ is equal to the number of shaper output bits, i.e. $n_{sh} = k_{sign} + \Delta n$.

For this procedure, coderate $R \in (0, 1]$ is fixed, while shaper input size $k_{sh}$ and shaper output $n_{sh}$ vary. The values of $k_{sign}$ and $n_{FEC}$ can be calculated using the code rate formulas. 

Extra bits are now shared between Shaper with rate $R_{sh} = \frac{k_{sh}}{n_{sh}}$ and Encoder with rate $R_{\textit{FEC}} = \frac{n_{sh} + k_{sign}}{n_{\textit{FEC}}}$, afterall $R = \frac{1}{2}(R_{sh} + 2R_{\textit{FEC}} -1)$. 

The problem is to find the optimal proportion between $R_{sh}$ and $R_{\textit{FEC}}$.



\subsection{Example of generated probabilities using ESS}

In Tabs.~\ref{tab:qam16_probs},~\ref{tab:qam64_probs} examples of generated probabilities for QAM-16 and QAM-64 (Fig.~\ref{fig:const_plot}) using the ESS method~\cite{gultekin2020probabilistic} are given. Note that for QAM-64 and above the probabilities of zeros and ones depend on each other, so joint probabilities need to be determined.

\begin{table}[]
\centering
\caption{Example of amplitude probabilities for QAM-16, $n_{sh}=256$.}

\begin{tabular}{|c|l|l|l|l|l|l|l|}
\hline
\textbf{$k_{sh}$} & 40   & 80   & 120  & 160  & 200  & 240 & 256 \\ \hline
\textbf{$p_a(0)$}   & 0.97 & 0.94 & 0.90 & 0.84 & 0.76 & 0.64 & 0.5 \\ \hline
\textbf{$p_a(1)$}   & 0.03 & 0.06 & 0.10 & 0.16 & 0.24 & 0.36 & 0.5 \\ \hline
\end{tabular}
\label{tab:qam16_probs}
\end{table}

\begin{table}[]
\centering
\caption{Example of amplitude probabilities for QAM-64, $n_{sh}=1024$.}

\begin{tabular}{|c|l|l|l|l|l|l|l|l|l|}
\hline
\textbf{$k_{sh}$}  & 300  & 400  & 500  & 600  & 700  & 800  & 900  & 1000 & 1024 \\ \hline
\textbf{$p_a(00)$}   & 0.87 & 0.80 & 0.73 & 0.66 & 0.59 & 0.52 & 0.44 & 0.32 & 0.25 \\ \hline
\textbf{$p_a(01)$}   & 0.13 & 0.18 & 0.23 & 0.27 & 0.30 & 0.32 & 0.31 & 0.28 & 0.25 \\ \hline
\textbf{$p_a(10)$}   & 0.00 & 0.01 & 0.03 & 0.06 & 0.09 & 0.13 & 0.18 & 0.23 & 0.25 \\ \hline
\textbf{$p_a(11)$}   & 0.00 & 0.00 & 0.00 & 0.01 & 0.01 & 0.03 & 0.07 & 0.16 & 0.25 \\ \hline
\end{tabular}
\label{tab:qam64_probs}
\end{table}

\subsection{Probability Shaping Mapping}

\begin{figure}
    \centering
    \includegraphics[width=1\linewidth]{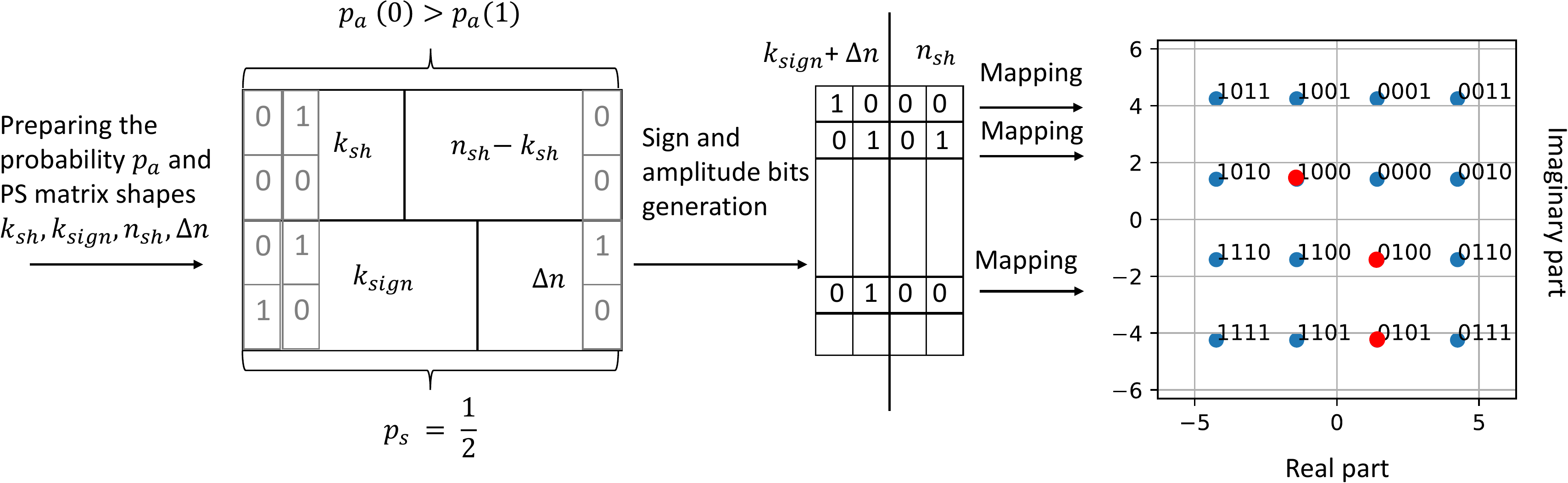}
    \caption{Creating a code block virtual matrix, generating sign $k_{sign}$ bits, LDPC $\Delta n$ bits and amplitude $n_{sh}$ bits with final symbol mapping. The red dots represent the mapped constellation points from the generated binary sequence.}
    \label{fig:ps_mapping}
\end{figure}

For mapping purposes, we form a special PS matrix (Fig.~\ref{fig:ps_mapping}) with uniform sign bits and non-uniform amplitude bits, following the data flow scheme (Fig.~\ref{fig:ps_scheme}). We generate $k_{sign}$ sign bits with equal probability of zeros and ones   $p_s=\frac{1}{2}$, and $n_{sh}$ amplitude bits with unequal probability of zeros and ones: $p_a \ne \frac{1}{2}$ such that $p_a(0) > p_a(1)$. The probability of amplitude bits can be determined by the proper values of $k_{sh}$ and $n_{sh}$ using the ESS method~\cite{gultekin2020probabilistic}. 




Finally, after the PS matrix is constructed, the mapping to the QAM is performed. With mapping procedure, bits are converted to the constellation points (or symbols) using the mapping table, which gives a specific coordinate on the complex plane for each unique sequence of bits. Notice that, given bit probabilities, there is a one-to-one correspondence to symbol probabilities.

After mapping procedure is done the symbols  go through the MIMO channel, demodulation and decoding, probability deshaping and BLER calculation, which are described in Sec.~\ref{sec:system_model} and Fig.~\ref{fig:pscm}. The demodulation and deshaping methods are the same procedures described earlier and are performed in reverse order. The decoding procedure is a complex process, which uses loopy belief propagation~\cite{murphy2013loopy} to iteratively recover the correct bits (LLRs).




\subsection{Arrangement of Finite Code Block Shapes}

To consistent all the shapes $k_{sh}$, $k_{sign}$ and $\Delta n$, and form the PS matrix (Fig.~\ref{fig:ps_mapping}) we solve the system of integer equations~\eqref{eq:integer system} finding LCM. Hereafter, the values of $N_{fr}^{sh}$ and $N_{fr}^{\textit{FEC}}$ define the multiplicative constants balancing these equations. The values of $N_A$ and $N_S$ define the number of amplitude and sign bits in the code block. It is implicitly assumed that everywhere in Figs.~\eqref{fig:ps_scheme}~\eqref{fig:ps_mapping} the values are $k_{sh} := N_{fr}^{sh} k_{sh}$, $k_{sign} := N_{fr}^{sh} k_{sign}$, $\Delta n := N_{fr}^{\textit{FEC}} \Delta n$, $n_{\textit{FEC}} := N_{fr}^{\textit{FEC}} n_{\textit{FEC}}$:
\begin{equation}
    \begin{cases}\label{eq:integer system}
    N_A = n_{sh} N_{fr}^{sh} \\ 
    CN_S = N_A \\
    N_S + N_A = N_{fr}^{\textit{FEC}} n_{\textit{FEC}},
    \end{cases}
\end{equation}
where the value of $C$ defines the constellation system, i.e. $C=1$~--- QAM-16, $C=2$~--- QAM-64, $C=3$~--- QAM-256 and so on. %

\section{Numerical Experiments}

\begin{figure}
    \centering
    \includegraphics[width=\linewidth]{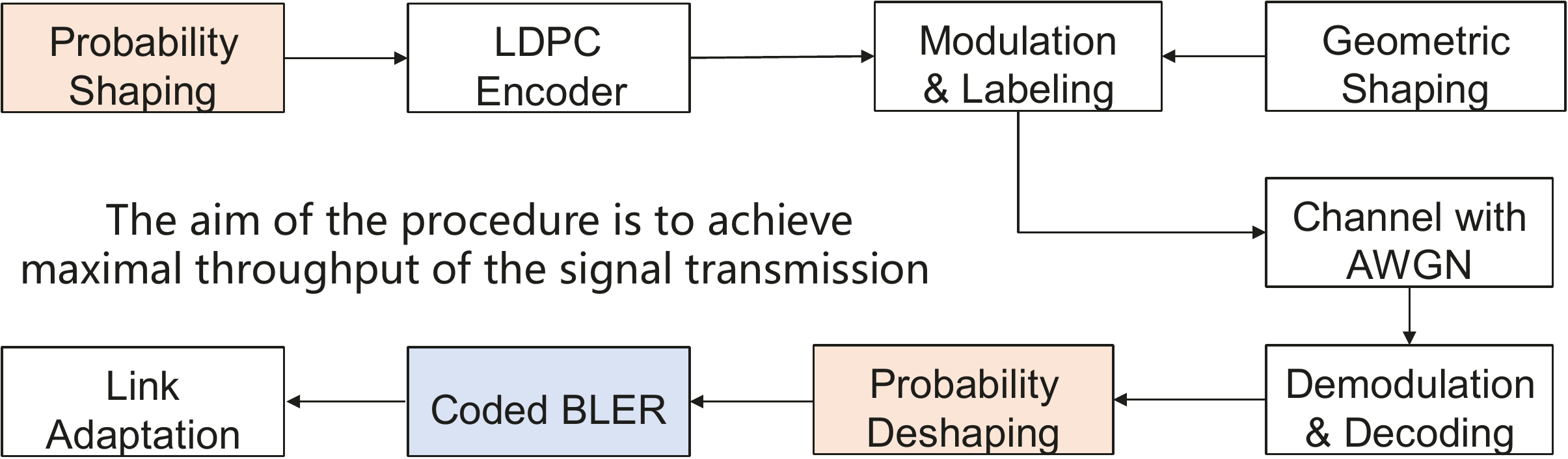}
    \caption{Block diagram of probability shaping transmitter and receiver.}
    \label{fig:pscm}
\end{figure}

In the experiments, Coded BLER is the average error of transmitted block of bits before the LDPC encoder and after the decoding in Fig.~\ref{fig:pscm}, which takes into account the realistic coding-encoding procedure.

\subsection{Energy per Bit and Noise Ratio}

The energy per bit to noise ratio $E_b/N_0$ is a normalized SNR measure, also known as SNR per bit. The $E_b/N_0$ measure can be used to express the relationship between signal power and noise power.

The energy per bit measure $E_b$ is the energy we use to transmit one bit of information with the total power $P$ and the LDPC coderate $R$:
\begin{equation*}
    E_b = \frac{P}{R},
\end{equation*}

The noise measure $N_0$ is the noise variance per real and imaginary parts:
\begin{equation*}
    N_0 = 2\sigma^2
\end{equation*}

Thus, $E_b/N_0$ can be expressed in terms of SNR:
\begin{equation*}
    E_b/N_0 = \frac{P}{R} \frac{1}{2\sigma^2} = \frac{P}{\sigma^2} \frac{1}{2R}  = \frac{\textrm{SNR}}{2R}
\end{equation*}

In decibel, $E_b/N_0$ is
\begin{equation}\label{eq:energy_bits_noise}
    E_b/N_0 \text{ in dB} = 10 \log_{10}(E_b/N_0) = 10 \log_{10}\left(\frac{P}{\sigma^2} \frac{1}{2R}\right)
\end{equation}

We use the value of $E_b/N_0$ in the Monte Carlo experiments. For a given value of $E_b/N_0$ with the power $P$ and coderate $R$ the variable noise power $\sigma^2$ disturbs the symbols transmitted over the channel~\eqref{eq:system}.

\subsection{Realistic Simulations using Sionna}

This study considers OFDM MIMO with a base station and a user equipped with multiple cross-polarised antennas. We provide simulations in Sionna~\cite{hoydis2022sionna} on the OFDM channel using 5G LDPC codes. The architecture of the system consists of LDPC, Bit Interleaver, Resource Grid Mapper, LS Channel Estimator, Nearest Neighbor Demapper, LMMSE Equalizer~\cite{mineev2023interference}, OFDM Modulator and presented in Fig.~\ref{fig:pscm}. Optimization variables are constellation type, a bit order, coderate, BLER, SNR, code block sizes and 5G model (LOS D, NLOS A).

The system uses soft estimates of LLRs for the decoder. Channel model is chosen to be OFDM 5G 2.6 GHz with delay spread of 40ns. The block size is 1536 with $10^5$ number of Monte-Carlo trials in the simulations and so in total $1.536 \cdot 10^8$ bits were processed for each point of $E_b/N_0$. For all simulation, 20 iterations of LDPC have been used. The code source is random binary tensors. The system use 3GPP wireless both Line of Sight (LOS) and  Non Line of Sight (NLOS) channel models D and A. The model is simulated in real time domain considering inter-symbol (IS) and inter-carrier (IC) interferences. In Tab.~\ref{tab:simulation_parameters} we provide simulation parameters for Sionna.

\begin{figure}
    \centering
    \includegraphics[width=1\linewidth]{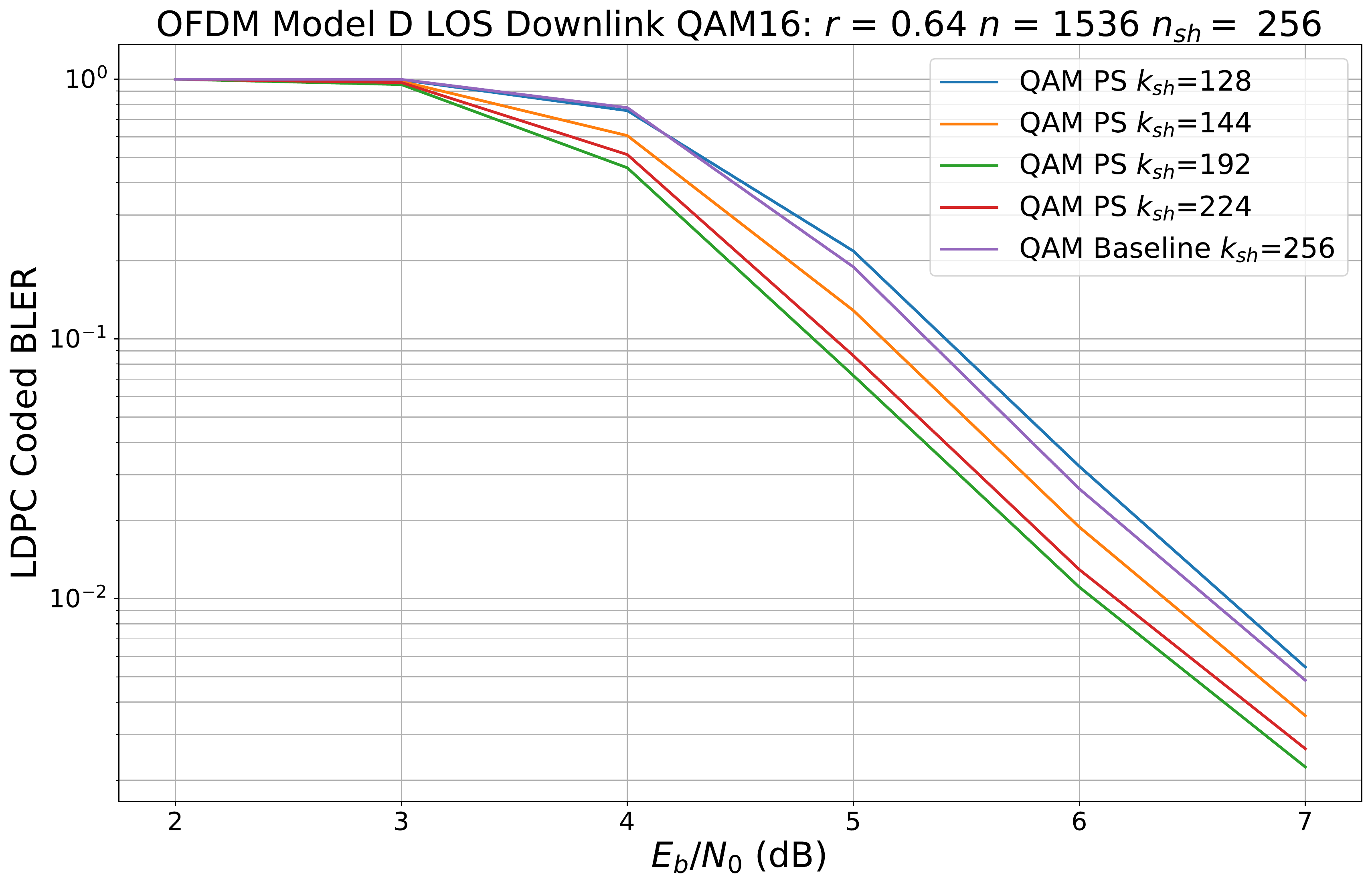}
    \caption{Model D LOS Downlink channel Coded BLock Error Rate for OFDM QAM16.}
    \label{fig:ofdm_coded_bler_los_downlink}
\end{figure}

\begin{figure}
    \centering
    \includegraphics[width=1\linewidth]{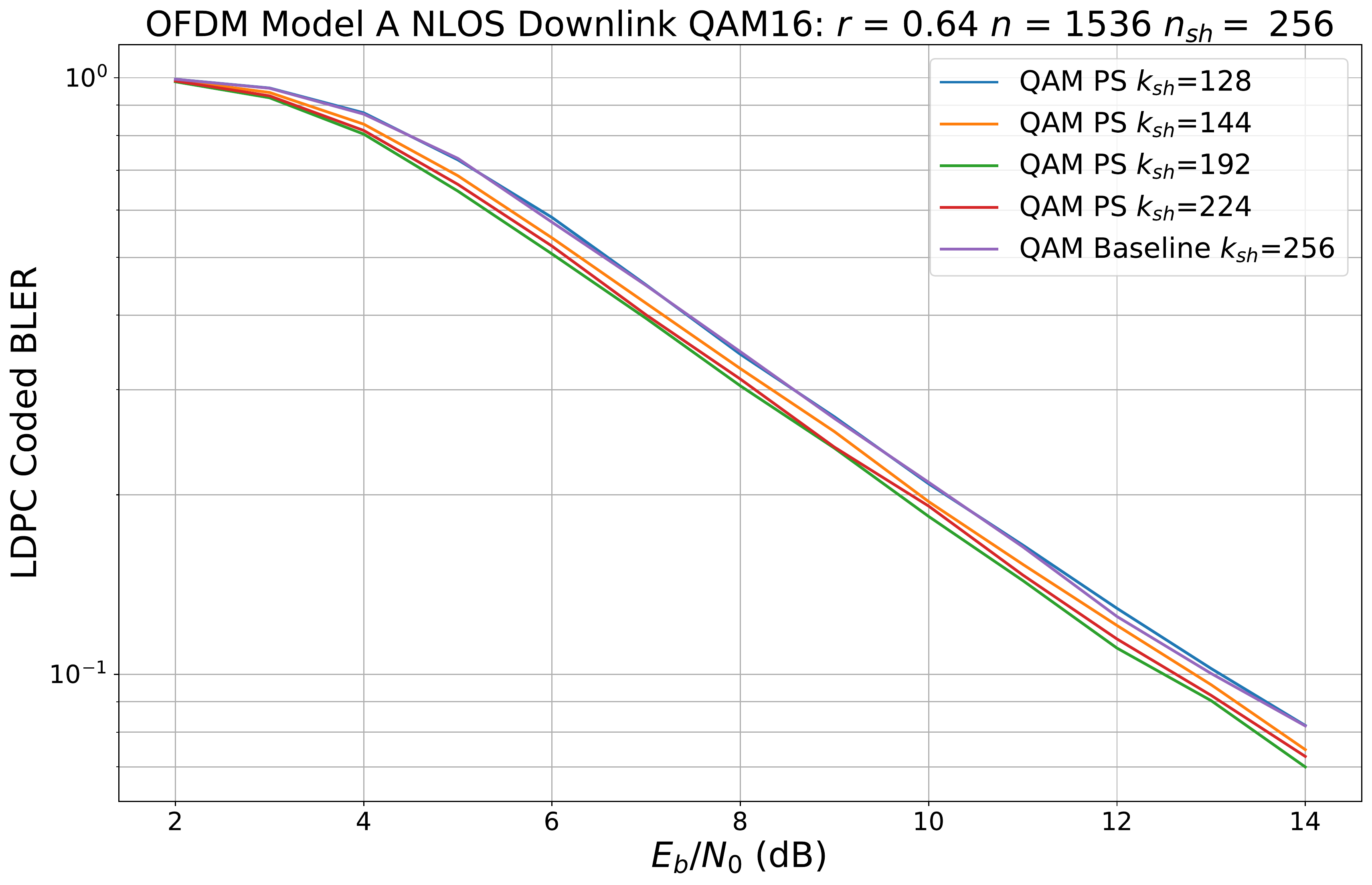}
    \caption{Model A NLOS Uplink channel Coded BLock Error Rate for OFDM QAM16.}
    \label{fig:ofdm_coded_bler_nlos_downlink}
\end{figure}

In Figs.~\ref{fig:ofdm_coded_bler_los_downlink} and~\ref{fig:ofdm_coded_bler_nlos_downlink}, we provide an experiment for both QAM-16 LOS Model D and NLOS Model A probability shaped (PS) constellations. We present experiments Coded BLER (see Fig.~\ref{fig:pscm}) with Gaussian transmit signal with QAM16 Baseline and amplitude PS with different shaping parameters,  where coderate is~$r$, block size is $n$ and parameters of PS are $n_{sh}$ and $k_{sh}$.

There is a local optimum for $k_{sh}=192$ PS QAM-16 in both LOS and NLOS models. It is noteworthy that the optimal parameter $k_{sh}$ is the same for the different LOS and NLOS models, which tells us that the chosen parameters are stable. Note that with wrong parameter settings, e.g. a strong shaping factor $k_{sh} = 128$, the quality of the PS method is worse than that of baseline QAM.

\begin{figure}
    \centering
    \includegraphics[width=1\linewidth]{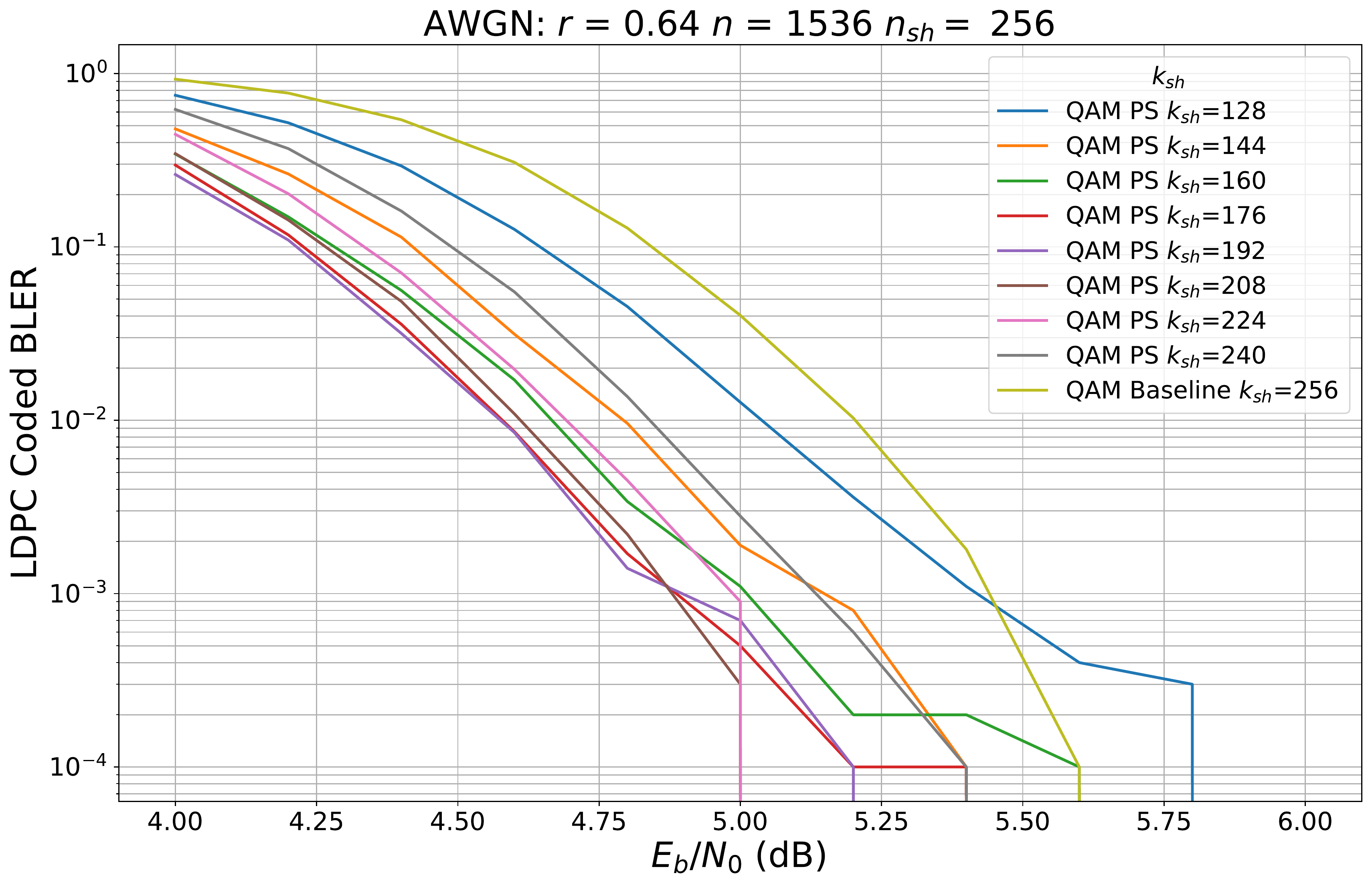}
    \caption{AWGN channel Coded BLock Error Rate QAM16.}
    \label{fig:awgn_coded_bler}
\end{figure}

In Fig.~\ref{fig:awgn_coded_bler} present experiments with AWGN channel and LDPC, which show a higher gain than for the OFDM channel model.

In Tab.~\ref{tab:gain_db_bler}, we provide gains in dB of the proposed PS method at 10\% BLER. The PS method achieves 0.56 dB gain in Model A NLOS Uplink compared to the baseline. From the massive experiments, we conclude that the proposed PS method constellations superior the baseline QAM for real 5G Wireless System using FEC LDPC.

The model codes and related experiments can be found in the repository\footnote{\href{https://github.com/eugenbobrov/On-Probabilistic-QAM-Shaping-for-5G-MIMO-Wireless-Channel-with-Realistic-LDPC-Codes}{https://github.com/eugenbobrov/On-Probabilistic-QAM-Shaping-for-5G-MIMO-Wireless-Channel-with-Realistic-LDPC-Codes}}.

\begin{table}
    \centering
    \caption{Simulation parameters in Sionna.}

    \begin{tabular}{|l|l|}
    \hline
    Carrier frequency    & 2.6e9      \\ \hline
    Delay spread         & 40e-9      \\ \hline
    Cyclic prefix length & 6          \\ \hline
    Num guard carriers   & {[}5, 6{]} \\ \hline
    FFT size             & 44         \\ \hline
    Num user terminal antennas           & 2          \\ \hline
    Num base station antennas           & 2          \\ \hline
    Num OFDM symbols     & 14         \\ \hline
    Num LDPC iterations  & 20 \\ \hline
    \end{tabular}
    \label{tab:simulation_parameters}
\end{table}

\begin{table}
    \centering
    \caption{Gain in dB of the proposed PS method at 10\% BLER.}

    \begin{tabular}{|l|c|c|}
    \hline
             & Model D LOS & Model A NLOS \\ \hline
    Uplink   & 0.52 dB     & 0.56 dB       \\ \hline
    Downlink & 0.5 dB     & 0.5 dB      \\ \hline
    \end{tabular}
    \label{tab:gain_db_bler}
\end{table}

\section{Conclusions and Suggested Future Work}

In this paper, for a MIMO OFDM wireless channel with realistic LDPC code at a given code rate, we study the PS scheme of Enumerative Sphere Shaping (ESS) known from the literature. We find local optimal parameters for the ESS method that minimise the BLER and provide a gain of up to 0.6 dB over the QAM-16 baseline through numerical experiments on the state-of-the-art Sionna simulation platform, modeling physical communication system level. Since there are almost no published works on PS that consider such realistic and contemporary scenarios, while only considering theoretical distributions, this study could be of scientific interest. In the future, a detailed study of BLER performance of a combination of PS and GS methods is possible, which could be very promising in communication applications.  

\section*{Acknowledgements}

The authors are grateful to Sergey Loktev, Dmitry Minenkov, Dmitry Shmelkin, Sviatoslav Panchenko, Alexandr Khodunin and Ivan Sobolev. 


\bibliographystyle{splncs04}
\bibliography{mybibliography}

\section*{Abbreviations}

\begin{tabular}{c|c}
    5G & Fifth Generation \\
    AWGN & Additive White Gaussian Noise \\
    BICM & Bit-interleaved Coded Modulation \\
    BLER & Block Error Rate \\
    CM & Coded Modulation \\ 
    ESS & Enumerative Sphere Shaping \\
    FEC & Forward Error Correction \\
    GS & Geometric Shaping \\
    IC & Inter-Carrier \\
    IS & Inter-Symbol \\
    LCM & Least Common Multiple \\
    LDPC & Low-Density Parity-Check Code \\
    LLR & Log-Likelihood Ratio \\
    LMMSE & Linear Minimum Mean Squared Error \\
    LOS & Line-of-Sight \\
    LS & Least Squares \\
    LTE & Long-Term Evolution \\
    MCS & Modulation and Coding Scheme \\
    MIMO & Multiple-input multiple-output  \\
    NLOS & Non-Line-of-Sight \\
    NUC & Non-uniform Constellations \\
    OFDM & Orthogonal Frequency-Division Multiplexing \\
    PS & Probability Shaping \\
    PSCM & Probability Shaped Coded Modulation \\
    QAM & Quadrature Amplitude Modulation \\
    SNR & Signal-to-Noise Ratio
\end{tabular}

\end{document}